\documentclass[amssymb, aps, prc, longbibliography, twocolumn]{revtex4-2}

\usepackage{graphicx}
\usepackage{amsmath}
\usepackage{multirow}
\usepackage{xcolor}
\usepackage{braket}

\begin{document}

\begin{abstract}
Excited states in  $^{13}$O were investigated using inelastic scattering of an E/A=69.5-MeV $^{13}$O beam off of a $^9$Be target. The excited states were identified in the invariant-mass spectra of the decay products. Both single proton and sequential two-proton decays of the excited states were examined. For a number of the excited states, the protons were emitted with strong anisotropy where emissions transverse to the beam axis are favored.  The measured proton-decay angular distributions were compared to predictions from distorted-wave born-approximation (DWBA) calculations of the spin alignment which was shown to be largely independent of the excitation mechanism. The deduced $^{13}$O level scheme is compared to \textit{ab initio} no-core shell model with continuum (NCSMC) predictions. The lowest-energy excited states decay isotropically  consistent with predictions of strong proton $1s_{1/2}$ structure.  
Above these states in the level scheme, we observed a number of higher-spin states not predicted within the model. Possibly these are associated with rotational bands built on deformed cluster configurations predicted by  antisymmetrized molecular dynamics (AMD) calculations. The spin alignment mechanism is shown to be useful for making spin assignments and may have widespread use.    

\end{abstract}

\title{Using spin alignment of inelastically-excited fast beams to make spin assignments: the spectroscopy of $^{13}$O as a test case}

\author{R.~J.~Charity}
\author{T.~B.~Webb}%
\author{J.~M.~Elson}
\author{D.~E.~M.~Hoff} \altaffiliation[Present Address: ]{Lawrence Livermore National Laboratory, Livermore, CA 94550}

\author{C.~D.~Pruitt} \altaffiliation[Present Address: ]{Lawrence Livermore National Laboratory, Livermore, CA 94550}

\author{L.~G.~Sobotka}
\affiliation{Departments of Chemistry and Physics, Washington University, St. Louis, Missouri 63130, USA}

\author{P.~Navr\'{a}til}
\affiliation{TRIUMF, 4004 Wesbrook Mall, Vancouver, British Columbia, V6T 2A3, Canada}
\author{G.~Hupin}
%\affiliation{Institut de Physique Nucl\'eaire, CNRS/IN2P3, Universit\'e Paris-Sud, Universit\'e Paris-Saclay, F-91406, Orsay, France}
\affiliation{Universit\'{e} Paris-Saclay, CNRS/IN2P3, IJCLab, 91405 Orsay, France}
\author{K.~Kravvaris}
\author{S.~Quaglioni}
\affiliation{Lawrence Livermore National Laboratory, P.O. Box 808, L-414, Livermore, California 94551, USA}

\author{K.~W.~Brown}
%\author{J.~Barney}
\author{G.~Cerizza}
\author{J.~Estee}
%\author{G.~Jhang}
\author{W.~G.~Lynch}
\author{J.~Manfredi} \altaffiliation[Present Address: ]{Air Force Institute of Technology, Wright-Patterson Air Force Base, Ohio 45433, USA}

\author{P.~Morfouace}
\author{C.~Santamaria}
\author{S.~Sweany}
\author{M.~B.~Tsang}
\author{T.~Tsang}
%\author{Y.~Zhang}
\author{K.~Zhu}
\affiliation{National Superconducting Cyclotron Laboratory, Michigan State University, East Lansing, Michigan 48824, USA}
\author{S.~A.~Kuvin} \altaffiliation[Present Address: ]{Los Alamos National Laboratory, Los Alamos, New Mexico 87545, USA}
\author{D.~McNeel}
\author{J.~Smith}
\author{A.~H.~Wuosmaa}
\affiliation{Department of Physics, University of Connecticut, Storrs, Connecticut 06269, USA}
\author{Z.~Chajecki}
\affiliation{Department of Physics, Western Michigan University, Kalamazoo, Michigan 49008, USA}
\date{present}%
\maketitle
%\tableofcontents

\section{INTRODUCTION}

A number of recent measurements have shown that the inelastic excitation of fast beams can result in excited states with strong spin alignment parallel and antiparallel to the beam axis \cite{Charity:2015,Hoff:2017,Hoff:2018,Charity:2018}. This spin alignment was observed only when the target nucleus remained in its ground state after the reactions. Hoff \textit{et al.} showed this alignment was due to a matching condition in grazing collisions and not dependent on the excitation mechanism but could be understood from general principles \cite{Hoff:2017,Hoff:2018}. For an excited state with the same parity as the beam, Hoff \textit{et al.} showed that this matching condition largely restricts collisions to those  where the  magnitude of the orbital angular momentum was unchanged. If the projectile's spin changes, angular momentum conservation is achieved by the tilting of the orbital angular momentum vector during the reaction.

While this spin alignment mechanism had previously been observed in inelastic excitations to states of known spin, its general applicability  may portend to widespread use in making spin assignments for newly observed states. In this work we will utilize this spin alignment mechanism to assign, or restrict, spins of $^{13}$O levels produced in inelastic scattering on $^{9}$Be target and identified with the invariant-mass method. 

The $Z$=8 closed shell  starts to quench as one approaches the proton drip line at $^{13}$O. Therefore, we expect to observe a  number of low-energy single and multiple $\hbar\Omega$ configurations. Some evidence for these has been observed in its mirror $^{13}$B \cite{Ota:2008,Iwasaki:2009,Guess:2009}, but the spectroscopy of $^{13}$O is quite sparse with only the ground state is particle bound. While \textit{ab initio} calculations of nuclear structure are restricted to light nuclei, most such calculations do not consider coupling to the continuum. The no-core shell-model (NCSM) approach \cite{Navratil:2000,Navratil:2000a,Barrett:2013} has been extensively applied to many particle-bound states in  light nuclei. It has recently 
 been extended to include coupling to the continuum \cite{Baroni:2013,Baroni:2013a,Navratil:2016}  and therefore the newly observed $^{13}$O excited states in this work provides a suitable test of this extension. 

A description of the experiment is presented in Sec.~\ref{sec:exp} and the experimental results are given in Sec.~\ref{sec:results}. The spin assignments of the observed levels are discussed in Sec.~\ref{sec:DWBA} and the theoretical calculations with the no-core shell model with continuum (NCSMC) are presented in Sec.~\ref{sec:NCSMC}. Section~\ref{sec:discussion} discusses the structure of the observed states with a comparison to the NCSMC predictions and finally the conclusions of this work are given in Sec.~\ref{sec:conclusions}.

\section{EXPERIMENTAL METHOD}
\label{sec:exp}
The data were obtained at the National Superconducting Cyclotron Laboratory at Michigan State University, which provided a ${}^{16}$O primary beam at 150 MeV/A. This primary beam bombarded a 193-mg/cm$^{2}$ ${}^{9}$Be target, and $^{13}$O fragments were selected by the A1900 magnetic separator. Upon extraction from the separator, this 69.5-MeV/A secondary beam had a purity of only 10\%. To remove the substantial contamination, the beam was sent into an electromagnetic time-of-flight filter, the Radio Frequency Fragment Separator \cite{bazin2009}, and emerged with a purity of 80\%. The final secondary beam impinged on a 1-mm-thick ${}^{9}$Be target and the charged particles produced were detected in the High Resolution Array (HiRA) \cite{WALLACE2007302} consisting of 14 $\Delta$\textit{E-E} [Si-CsI(Tl)] telescopes 85 cm downstream of the target. The array subtended a polar angular range of 2.1$^{\circ}$ to 12.4$^{\circ}$. Each telescope consisted of a 1.5-mm-thick, double-sided Si strip $\Delta E$ detector followed by a 4-cm-thick, CsI(Tl) $E$ detector. The $\Delta E$ detectors are 6.4~cm $\times$ 6.4~cm in area, with each of the faces divided into 32 strips. Each $E$ detector consisted of four separate CsI(Tl) elements each spanning a quadrant of the preceding Si detector. Signals produced in the 896 Si strips were processed with the HINP16C chip electronics \cite{Engel07}.

The energy calibration of the Si detectors was obtained with a $^{232}$U 
$\alpha$-particle source. The particle-dependent energy calibrations of the CsI(Tl)
detectors were achieved with cocktail beams selected with the A1900 separator.
Data from this experiment pertaining to the formation of $^{11}$O, $^{11}$N, $^{12}$O, $^{12}$N, and $^{13}$F resonances has already been published in Refs.~\cite{Webb:2019,Webb:2019a,Webb:2020,Charity:2020,Charity:2021}.

\section{RESULTS}
\begin{figure}
%sortcode_addback/tree/O13/invMass.C
\includegraphics[scale=0.4]{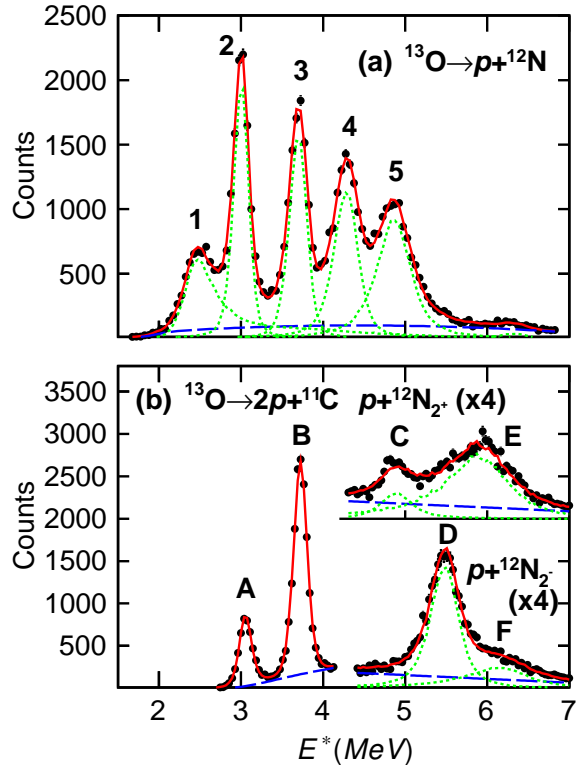}
\caption{Excitation-energy spectra for transverse decay determined with the invariant-mass method for $^{13}$O excited states produced in the inelastic scattering of an $^{13}$O beam. (a) shows the distribution obtained for the $p$+$^{12}$N exit channel while (b) shows the distribution for the 2$p$+$^{11}$C exit channel. For the latter, events with excitation energies above 4.3~MeV have been subdivided into those associated with the 2$^+$ and 2$^-$ intermediate states in $^{12}$N.  The results for the 2$^+$ intermediate state have been shifted up the $y$ axis for clarity. The solid-red curves show fits to these spectra with contributions from individual levels shown by the dotted-green curves and the background contributions by the dashed-blue curves.  The fitted levels are identified with a label (see Table~\ref{tbl:level}).}
\label{fig:invMass}
\end{figure}

\
\label{sec:results}
Distributions of the excitation energy obtained with the invariant-mass method for the $p$+$^{12}$N and 2$p$+$^{11}$C channels are shown in Figs.~\ref{fig:invMass}(a) and \ref{fig:invMass}(b), respectively. The best invariant-mass resolution is obtained by only including events that decayed transversely, i.e. the heavy core fragment is emitted transversely to the beam axis from the moving parent $^{13}$O$^*$ fragment \cite{Charity:2019,Charity:2019a}. The transverse gate used to produce these spectra is $\left|\cos\theta_{C}\right|<0.2$ where $\theta_{C}$ is the emission angle of the core from the beam axis in the parent's center-of-mass frame. 

In the $p$+$^{12}$N spectrum in Fig.~\ref{fig:invMass}(a), we see five peaks associated with levels which are referred to in the subsequent discussions by the labels \textit{1} to \textit{5} in ascending order of excitation energy.
 In the 2$p$+$^{11}$C spectrum one can also separate a similar number of peaks, but here they are associated with sequential two-proton decay through either the 2$^+$ first excited state of $^{12}$N or the 2$^-$ second excited state. To help separate the peaks at the higher excitation energies ($E^* > 4.3$~MeV) where they overlap, we have put gates on  these intermediate states, via the $p$+$^{11}$C invariant mass, producing two spectra.  The levels associated with these observed peaks are labeled \textit{A} to \textit{F} again in ascending order of excitation. 
Peaks \textit{A} and \textit{B} were observed in \cite{Sobotka:2013} and were shown to undergo sequentially proton decays through the 2$^+$ and 2$^-$ intermediate states of $^{12}$N, respectively. 
    
\begin{figure*}
%sortcode_addback/tree/O13/levelWithTheory2.C
\includegraphics[scale=0.8]{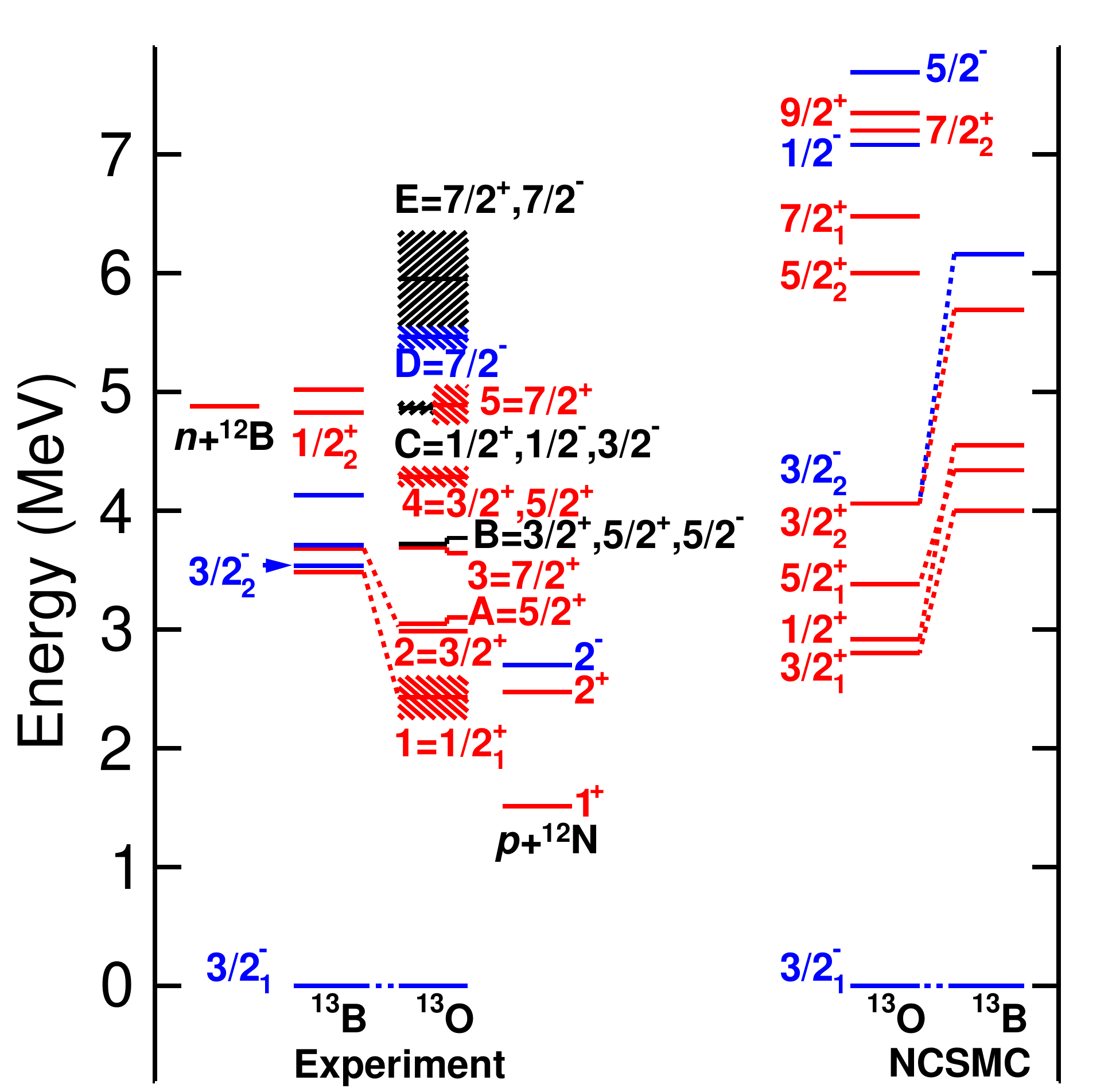}
\caption{Level diagram of $^{13}$O showing the levels identified in this work. For comparison, the experimental levels in the mirror nucleus $^{13}$B \cite{Back:2010} are also shown. In addition, on the right are level predictions from the NCSMC theory. Negative-parity states are show in blue while positive-parity states are in red. Levels where the parity is not determined are shown in black. Only the widths of the experimental levels are shown.  The predicted $^{13}$O states have wider widths which overlap.}
\label{fig:levelScheme}
\end{figure*}

The invariant-mass spectra were fit to extract centroids and decay widths of the observed levels, which are listed in Table~\ref{tbl:level} and a level scheme is shown in Fig.~\ref{fig:levelScheme}. The fits assumed intrinsic Breit-Wigner line shapes for the narrower levels with the experimental resolution incorporated via Monte-Carlo simulations \cite{Charity:2019}. The simulated experimental resolution at the centroid of peak \textit{2} is  230~keV FWHM, while for peak \textit{B}, the resolution is 180~keV.  For the wider levels (labeled \textit{1}, \textit{5}, \textit{E}, and \textit{F}), we have used intrinsic lines shapes from the $R$-matrix formalism \cite{Lane:1958} and the listed centroids and widths are given as the poles of the $S$-matrix for these resonances. In Fig.~\ref{fig:invMass}, the fitted distributions (solid-red curves) are the sum of the individual peaks (dotted-green curves) plus a smooth background (dashed-blue curves) which represents the contributions from non-resonance breakup and wide unresolved resonances.

\begin{table*}

\caption{Levels of $^{13}$O extracted in this work with the fitted excitation energy $E^*$,  decay width $\Gamma$, spin assignments $J^\pi$, and labels from Fig.~\ref{fig:invMass}.  The proton decay energies $E_{p}$ for decay to the ground and excited states $^{12}$N are also listed. Only the statistical error on the centroids are listed, but there is an additional systematic uncertainty of 8~keV. The spins of the $^{12}$N daughters following proton emission are also listed.}
\label{tbl:level}
\begin{ruledtabular}

\begin{tabular}{c c c c c c}
$E^*$ & $E_{p}$ & $\Gamma$ & $^{12}$N daughter & $J^\pi$ & peak label \\
(MeV) & (MeV) & (keV) \\
\hline
2.428(12)& 0.916(12) & 358(19) & 1$^+$ & $1/2^{+}$ & \textit{1} \\
3.006(1) & 1.494(1)  & 55(19)   & 1$^+$ & $3/2^{+}$ & \textit{2} \\
3.051(7) & 0.578(8) & 54(19)  & 2$^+$ & $5/2^{+}$  & \textit{A} \\
3.692(2) & 2.180(2) & 53(21)  & 1$^+$ &  3/2$^+$, 5/2$^+$, 5/2$^-$ & \textit{3} \\
3.721(1) & 1.019(7) &  10(19)  & 2$^-$ &  3/2$^+$, 5/2$^+$, 5/2$^-$ & \textit{B} \\
4.287(4) & 2.775(4) & 170(25) & 1$^+$ &  3/2$^+$, 5/2$^+$     & \textit{4} \\
4.866(12) & 1.293(14) &  103(37) & 2$^+$ &  1/2$^{+}$, 1/2$^{-}$, 3/2$^-$ & \textit{C} \\
4.892(22) & 3.380(22)& 323(27) & 1$^+$ &   7/2$^+$     & \textit{5} \\
5.483(5)  & 2.765(8) & 204(41) & 2$^-$ &   7/2$^-$     & \textit{D} \\
%5.467(5)  & 3.955(5) & 204(42) & 2$^-$ &  7/2$^-$    & \textit{D} \\

5.951(10) & 3.478(12) & 875(68) & 2$^+$&   7/2$^+$, 7/2$^-$ & \textit{E}\\
$\approx$6.2 &  $\approx$ 5.0        &         & 2$^-$ &       & \textit{F}\\ 
\end{tabular}

\end{ruledtabular}
\end{table*}

The peak labeled \textit{F} in Fig.~\ref{fig:invMass}(b) sits under the high-energy tail of the more intense peak \textit{D} and is not fully resolved. Without including this peak, the background would need to be peaked in this region to obtain a good fit, so its presence is significant, but its exact location and width is not well constrained. In the fit in Fig.~\ref{fig:invMass}(b), we have forced its width to have the same value as for peak \textit{E} in the $^{12}$N$_{2^{+}}$-gated spectrum with the assumption that peaks \textit{E} and \textit{F} could represent two decay branches of the same level. However, the best fit occurs when the centroid of \textit{F} is $\approx$250~keV higher in energy suggesting it may represent a separate level.  Due to the difficulties in constraining this peak, we will not consider it further.

\subsection{Comparison to previous studies}
\label{sec:previous}
From an $R$-matrix analysis of $p$+$^{12}$N resonant elastic scattering using the inverse-kinematics, thick-target technique, a $J^\pi$=1/2$^+$ first excited state was found at $E^*$=2.69(5)~MeV with a decay width of 350-550~keV \cite{Skorodumov:2007}.  The width is consistent with our peak \textit{1}, but the centroid is $\approx$260~keV higher in energy. For the latter, this could be partly due to different definitions of the resonance energy which diverge as a state becomes wider. This previous work also suggested a 3.29(5)-MeV level with spin of either 1/2$^-$ or 3/2$^-$ and width $\approx$75~keV. We do not see any evidence for this state.

Peaks at $E^*$=2.75(4), 4.21, and 6.03(8)~MeV have been observed in the $^{13}$C($\pi^{+}$,$\pi^{-}$) reaction \cite{Seidl:1984,Ward:1993}. The 6.02-MeV state is quite wide with a FWHM of 1.2~MeV and is possibly associated with peaks \textit{E} and \textit{F} in the present work. The 4.21-MeV peak in that work is most consistent with peak \textit{4} of this work.

In the $^{14}$O($p$,$d$) reaction, Suzuki \cite{Suzuki:2012} observed excited states at $E^*$=2.8(3) and 4.2(3)~MeV which decayed to the ground state of $^{12}$N. These are consistent with peaks \textit{2} and \textit{4} in this work.  From the angular distribution of the outgoing deuterons, the 4.2(3)-MeV  state was assigned a spin of either 1/2$^-$, 3/2$^{-}$, 3/2$^{+}$, or 5/2$^+$.

A lower-energy ($E/A$=30.3~MeV) inelastic-scattering study was also performed with the invariant-mass method \cite{Sobotka:2013}, but the statistics were significantly reduced compared to the present work.  Peaks \textit{2}, \textit{A}, and \textit{B} were observed in that study but at 20-50 keV lower in excitation energy. That study also discussed whether peaks \textit{2} and \textit{A} could be decay branches of the same state. With the higher statistics of this work we determine their decay widths are consistent and the shift between the centroids is only 45(7)~keV.  Small shifts between the centroids of two decay branches are possible due to the different energy dependencies of the barrier penetration factors across the resonance, but should be much less than the intrinsic width of the state. However in this case, the shift is approximately equal to the extracted decay widths of $\approx$55~keV for the two peaks. Also the relative yield of the two states determined in the lower-energy experiment was 29(8)\% (peak \textit{2}) and 71(8)\% (peak \textit{A}), significantly different from values of 51(3)\% and 49(3)\% respectively we obtain in the present study after correcting for the detection efficiency. This strongly suggests that  these two peaks arise from two separate states. Peaks \textit{3} and \textit{B} are separated by only 29(2)~keV, but this shift is also large compared to their extracted widths, thus suggesting these two are associated with separate levels as well.

\subsection{Decay Angular Distributions}
\begin{figure}
%sortcode_addback/tree/O13/Ex_target.C
\includegraphics[scale=0.4]{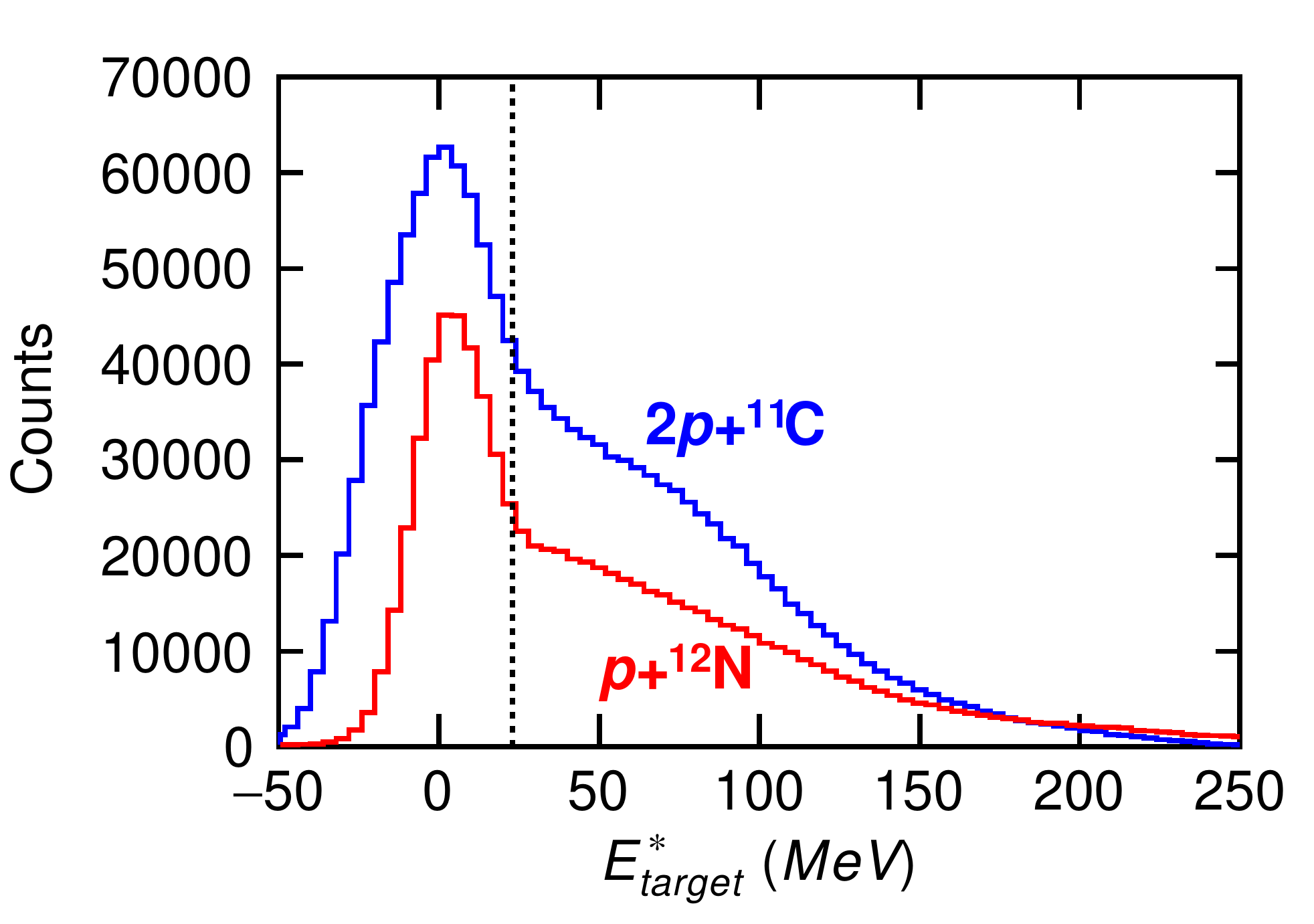}
\caption{Distribution of the excitation energy in the target nucleus after the inelastic scattering reaction obtained for both the $p$+$^{12}$N and 2$p$+$^{11}$C events. The dotted vertical line delimits ground-state and excited target nuclei in the subsequent analysis.} 
\label{fig:Ex_target}
\end{figure}

Previous studies of inelastic scattering with fast beams have shown the possibility of achieving strong spin alignment if the target nucleus remains in its ground state after the inelastic scattering \cite{Charity:2015,Hoff:2017,Hoff:2018,Charity:2018}. Figure \ref{fig:Ex_target} shows the distributions of target excitation energy deduced from the center of mass of the $p$+$^{12}$N and 2$p$+$^{11}$C events assuming two-body kinematics for inelastic scattering reactions. Both distributions display a prominent peak at $E^{*}_{target}$=0~MeV, the width of which is consistent with the expected resolution. The distributions extend to hundreds of MeV, which is very substantial for such a light nucleus. It not clear whether our assumption of two-body kinematics is correct for these events  and possibly other reaction mechanisms contribute to the high-energy tail. The dotted vertical line in this figure shows the delimiting value of excitation energy used to separate ground-state and excited target nuclei in the subsequent analysis.

\begin{figure}
%sortcode_addback/tree/O13/p12N/ang.C
\includegraphics[scale=0.45]{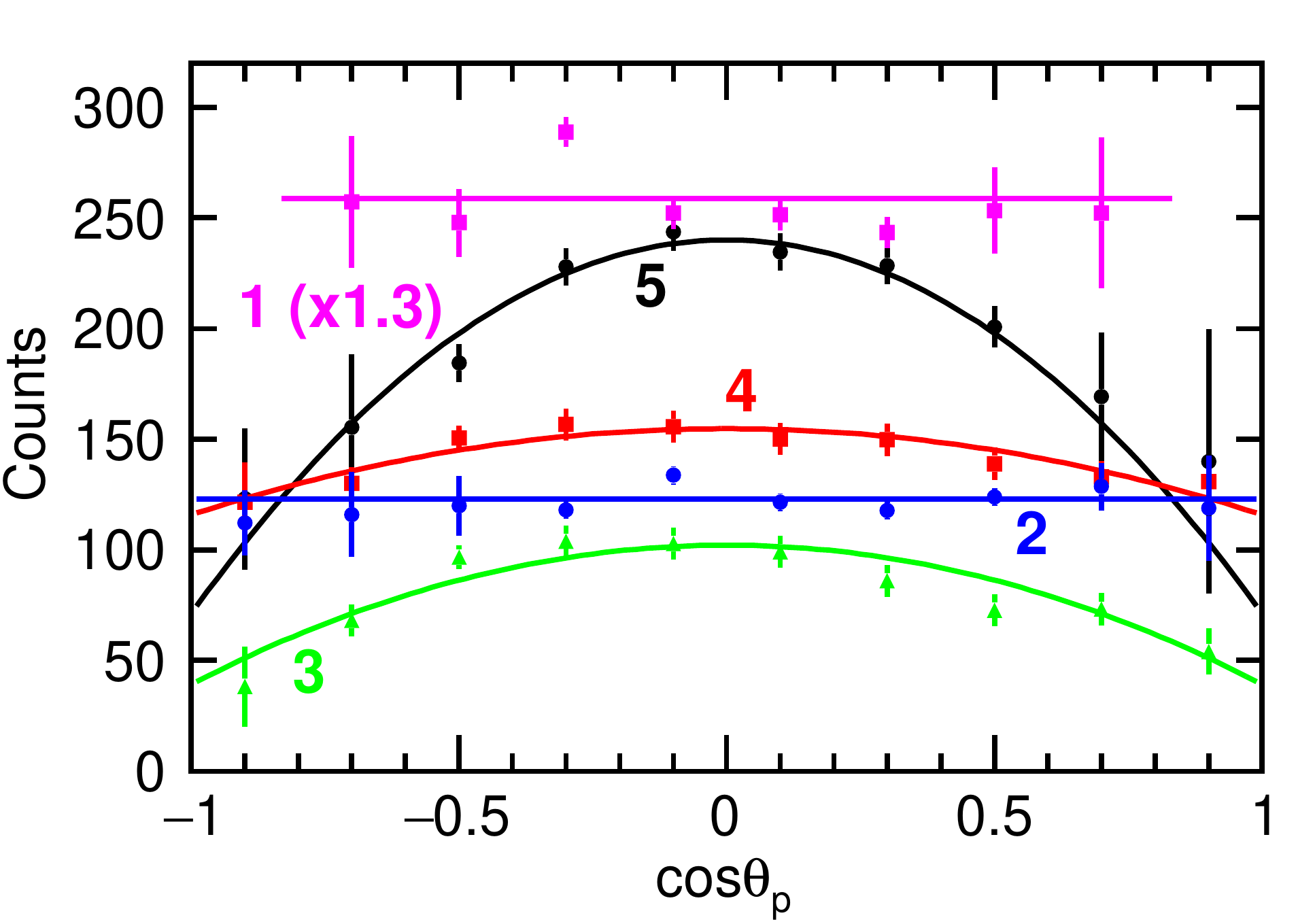}
\caption{Angular distribution of the decay protons in the parent $^{13}$O fragment's reference frame where $\theta_p$=0 corresponds to the proton being emitted along the beam axis. These distributions are for the peaks observed in the $p$+$^{12}$N exit channel were the target fragment is not excited. See Fig.~\ref{fig:invMass} and Table~\ref{tbl:level} for level labels. All distributions has been corrected for the detection efficiency. The lines guide the eye.}
\label{fig:ang1p}
\end{figure}

The angular distributions of the decay protons in the parent $^{13}$O$^*$ center-of-mass frame for peaks \textit{1} to \textit{5} are shown in Fig.~\ref{fig:ang1p} for events where the target fragment remained in its ground state.  These distributions have been corrected for the detection efficiency determined with Monte Carlo simulations \cite{Charity:2019}. Forward emission parallel to the beam axis corresponds to $\theta_p$=0$^\circ$.

\begin{figure}
%sortcode_addback/tree/O13/2p11C/ang.C
\includegraphics[scale=0.4]{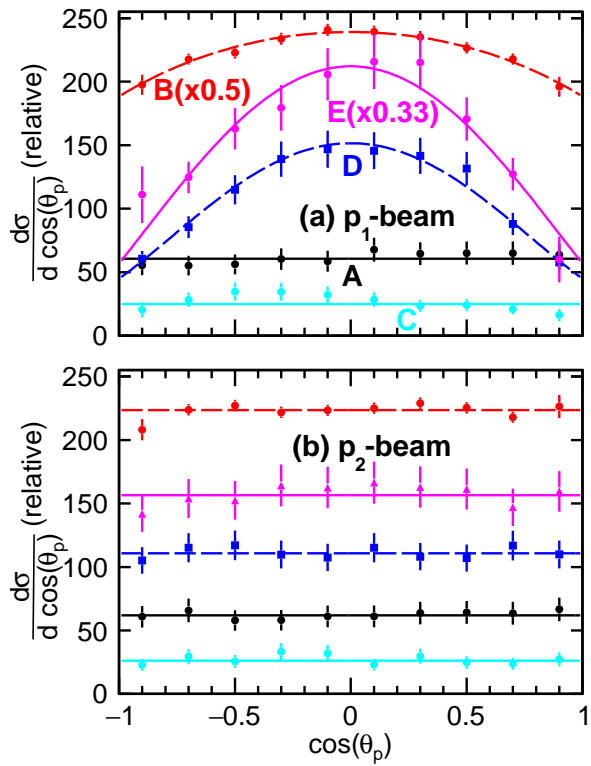}
\caption{Angular distributions for protons produced in the 2$p$ decay of $^{13}$O excited states where the target nucleus is not excited. The angular distributions in the moving $^{13}$O$^*$  reference frame for the first and second emitted protons are shown in panels (a) and (b), respectively. All distributions have been corrected for the detection efficiency.  For (a), curves show polynomial fits to these distributions enforcing symmetry about cos$\theta_P$=0. The solid and dashed curves differentiate levels which decay through the 2$^+$ and 2$^-$ intermediate states of $^{12}$N, respectively. For (b), the horizontal lines show fits assuming isotropic emission.}
\label{fig:ang2p}
\end{figure}

The peaks observed in the 2$p$+$^{11}$C channel correspond to sequential two-proton decay through $^{12}$N intermediate states. The emission energies of the two protons are different enough that these can be separated. See Ref.~\cite{Sobotka:2013} for a  discussion of the correlations between the protons for peaks \textit{A} and \textit{B} and the identification of the first and second emitted proton.  The angular distributions of the first and second  emitted proton for  peaks \textit{A}-\textit{E} are displayed in Fig.~\ref{fig:ang2p}(a) and \ref{fig:ang2p}(b), respectively,  for events where the target nucleus remains in its ground state. 

\begin{figure}
%sortcode_addback/tree/O13/2p11C/ang_nel_one.C
\includegraphics[scale=0.4]{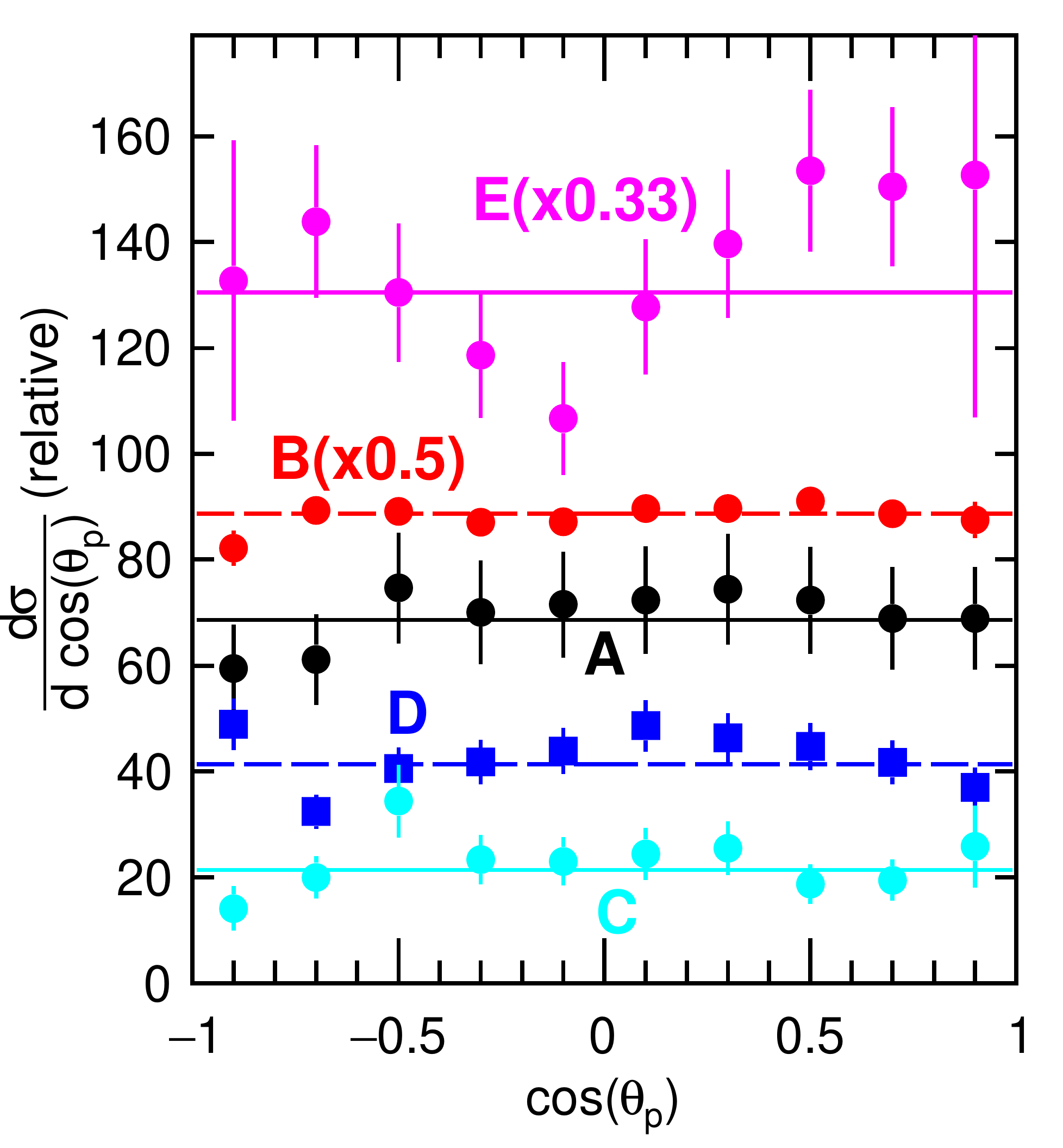}
\caption{Angular distributions for the first proton in the moving $^{13}$O$^*$ reference frame for states which 2$p$ decay and  where the target nucleus is  excited. The distributions have been corrected for detector efficiency and the horizontal lines show fits assuming isotropic emission.}
\label{fig:ang_nel}
\end{figure}

 The distributions of the second-emitted proton in Fig.~\ref{fig:ang2p}(b) are all consistent with isotropic emission. The 2$^+$ intermediate state is expected to decay by $s$-wave proton emission.  Even if there is some small $d$-wave admixture in its wavefunction, the barrier penetration factor at a decay energy of  
360~keV will suppress  $d$-wave decay. Thus we expect the observed isotropic decay. 

 The 2$^-$ intermediate state can proton decay to the ground state of $^{11}$C by $p_{1/2}$ and/or $p_{3/2}$ proton emissions. Pure $p_{1/2}$ decay is trivially isotropic, but we find also that pure $p_{3/2}$ decay in this particular case is also isotropic. The only source of any anisotropy would be from an interference between $p_{1/2}$ and $p_{3/2}$ decay components and would be observed in the angular distribution only if there was a strong spin alignment of this intermediate state.  For peak \textit{D} for which we inferred  strong spin alignment in Sec.~\ref{sec:DWBA}, the spin alignment of the 2$^-$ intermediate state after proton emission was found to be significantly reduced. The maximum deviation from isotropy is for  strong interference between the $p_{1/2}$ and $p_{3/2}$ emissions where the proton angular distribution which is flat within 12\%. This is consistent with the experimental data.

In contrast to the second-emitted proton, the angular distributions for 
the first-emitted proton  show a range of behaviors with some consistent with isotropic emission (peaks \textit{1}, \textit{2}, \textit{A}, and \textit{C}) and some with very strong anisotropies (peaks \textit{3}, \textit{5}, \textit{D} and \textit{E}). While  proton emission associated with events where the target nucleus remains in its ground state can have significant anisotropies in their angular  distributions, those associated with excited target nuclei have isotropic or near-isotropic angular distributions. For example, the angular distributions of the first protons for peaks \textit{A} to \textit{D} are shown in Fig.~\ref{fig:ang_nel} where they are all reasonably fit with flat, isotropic dependencies. The results were found to be similar for the peaks in the $p$+$^{12}$N spectrum with $E^*_{target}>0$.
These angular distributions indicate that the $E^*_{target}>0$ events are not associated with any significant spin alignment.
A similar dependence of the  spin alignment on $E^*_{target}$ for  fast $^7$Be projectiles was observed in Ref.~\cite{Charity:2015}.

\subsection{Dependence of Yields on Target Excitation Energy}

\begin{figure}
%sortcode_addback/tree/O13/Ex_el.C
\includegraphics[scale=0.4]{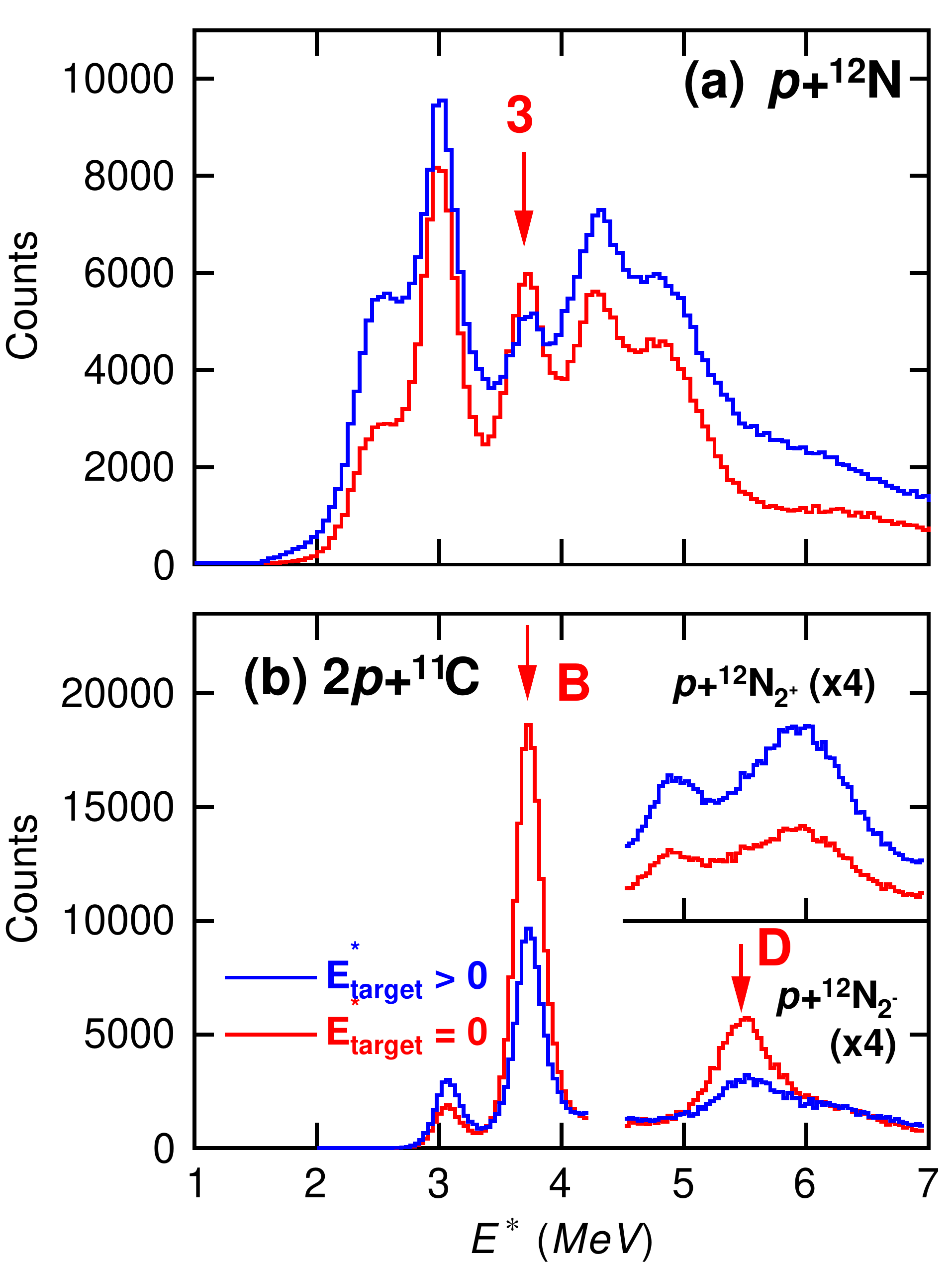}
\caption{Comparison of $^{13}$O excitation-energy distributions obtained with the invariant-mass technique where the target nucleus remains in its ground state (blue histograms) or is excited (red histograms). Otherwise the gating condition as the same as in Fig.~\ref{fig:invMass} except that there is now no gate on transverse decay in order to compare total yields.}
\label{fig:Ex_el}
\end{figure}

\begin{figure}
%sortcode_addback/tree/O13/suppress.C
\includegraphics[scale=0.4]{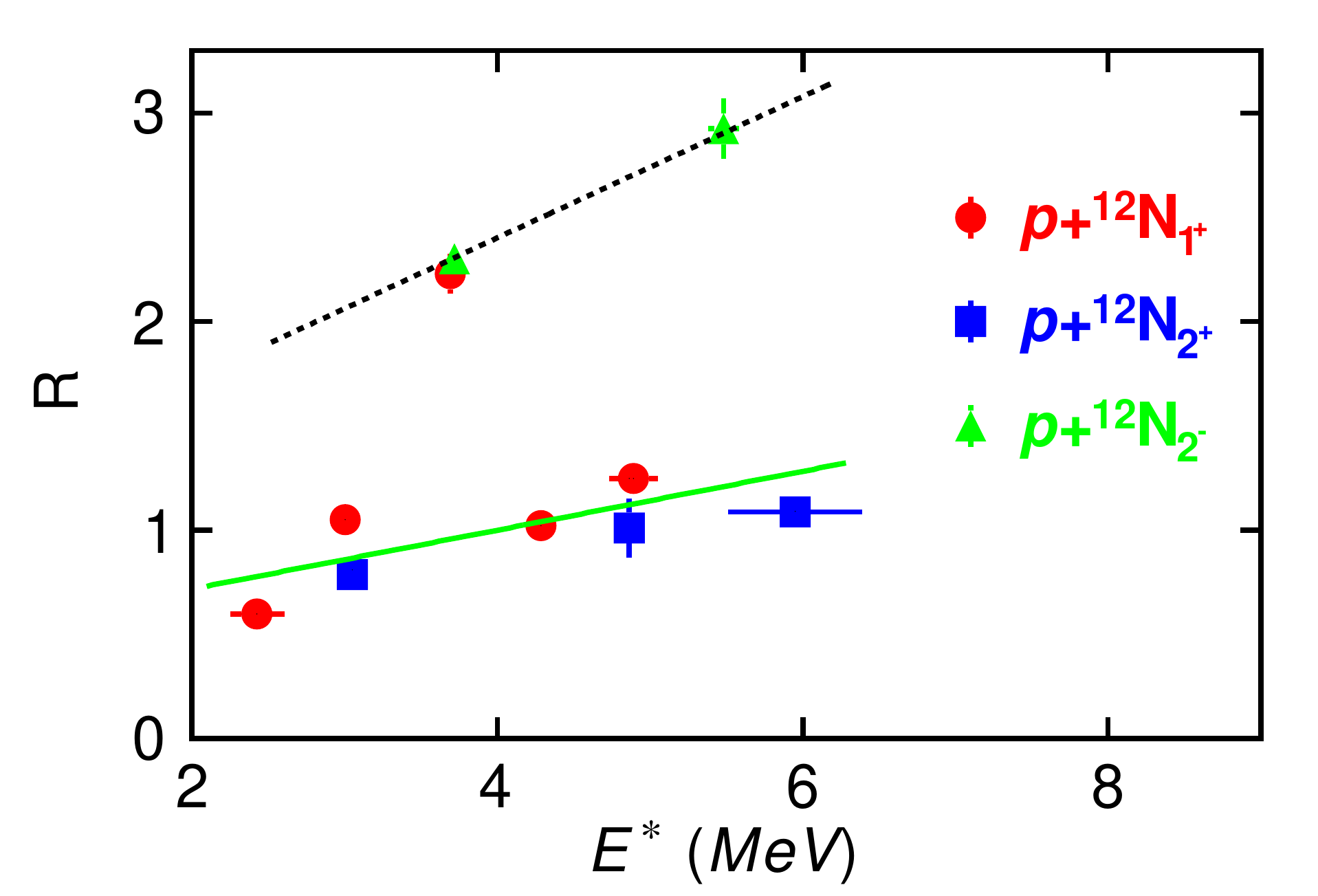}
\caption{
The  ratio  $R$ of $E^*_{target}=0$ to $E^*_{target}>0$  yields  for the observed $^{13}$O states  plotted as a function of the projectile excitation energy.
 The yields were obtained from fits to the histograms in Fig.~\ref{fig:Ex_el} corrected for detector efficiency.  The horizontal error bars are the fitted widths of the states.}
\label{fig:suppress}
\end{figure}

When subdividing the detected events  according to $E^*_{target}$, we found that the observed peaks could be cleanly separated into two groups.   Figure \ref{fig:Ex_el} compares the $^{13}$O invariant-mass distributions gated on the  ground-state (red histograms) and excited (blue histograms) target nuclei. Only for peaks \textit{3}, \textit{B}, and \textit{D} are the yields stronger for the E$^*_{target}>0$ gate. To make this more quantitative, peak yields for the two gates were fit, corrected for detector efficiency, and the ratio $R$ 
of events with $E^{*}_{target}=0$ to those with $E^{*}_{target}>0$ is plotted verses projectile excitation energy in Fig.~\ref{fig:suppress}. Most peaks have $R\approx$1.0 while the aforementioned peaks have $R>2$. For the peaks with the low $R$ values, there is a small dependence on the projectile excitation energy which was fit by the solid green line.   The black dotted line in Fig.~\ref{fig:suppress} is the same dependence scaled by a factor of 2.4 and reproduces the dependence of the remaining levels.  The separation between these two groups is remarkably clean. While a full understanding of this result is beyond this work given that there are uncertainties to the reaction mechanisms populating the higher $E^*_{target}$ values, it does suggest that the states with similar $R$ values have similar structure.

\section{DWBA calculations and spin assignments}
\label{sec:DWBA}
The peaks associated with the anisotropic proton decays must have spins greater then $J$=1/2.  Further restrictions to their spins can be obtained from knowledge of their spin alignment in the reactions. In the excitation of $^7$Li and $^7$Be projectiles from their 3/2$^-$ ground state to the 7/2$^-$ excited state, large spin alignments were found when the target nucleus remained in its ground state and it was argued that this was largely independent of the excitation mechanism but was based on general arguments \cite{Hoff:2017,Hoff:2018}.   In this work, we assume the proton decays from the observed levels can only be $s$, $p$, and $d$-wave in nature giving us a maximum possible spin of 9/2$^+$ for  decay to the excited states of $^{12}$N.   We assume that spin-dependent forces which can induce spin-flips  are not very important and the reactions are associated mainly with normal parity transitions induced by central forces \cite{ThompsonBook}, i.e, the parity changes for odd multipolarity transitions and remains unchanged for even multipolarities. The spin of the target nucleus was shown to be unimportant in Ref.~\cite{Hoff:2018} where similar spin alignments for $^7$Li excited state was observed with both a $J$=3/2 $^9$Be and a spin-zero $^{12}$C target. 

\begin{figure}
%sortcode_addback/tree/O13/fresco_mdist.C
\includegraphics[scale=0.43]{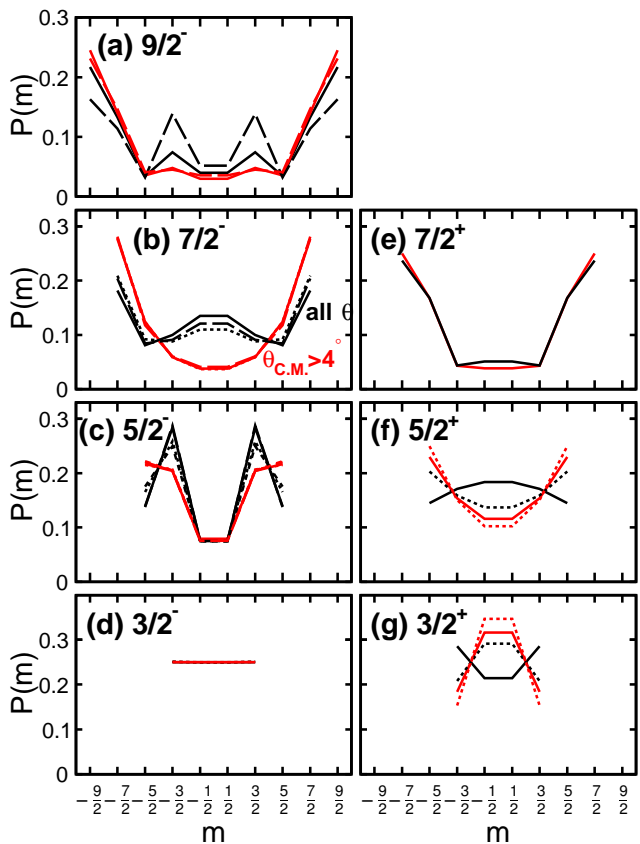}
\caption{Distributions of $m$-states calculated with the FRESCO code for different assumed spins and parities of the excited projectile.
The red lines show the predictions when $^{13}$O$^*$  center-of-mass scattering angles  less than 4$^\circ$ are excluded, while the black lines show the predictions integrated over all  angles. The solid lines show the results where the multipolarity of the excitation is specified, the long-dashed lines are from the rotor model, while the short-dashed lines are associated with single-particle proton and neutron excitations.}
\label{fig:fresco_mdist}
\end{figure}

 We have made distorted-wave Born approximation (DWBA) calculations with the code FRESCO \cite{Thompson:1988} assuming both single-particle excitations, a rotational model, and specifying the reduced matrix element for a particular multipolarity $K$ \cite{ThompsonBook}. The resulting $m$-state probability distributions $P(m)$, using the beam axis as the quantization direction, are shown in Fig.~\ref{fig:fresco_mdist} for possible spins and parities of the excited projectile. Due to the choice of the quantization axis, these distributions are symmetric about $m$=0 \cite{Charity:2015}. The $m$-state distributions show little dependence on the assumed reaction mechanism and the parameters used such as in the various optical-model potentials and the coupling parameters. This observation  confirms that the $P(m)$ distributions  are defined by general arguments.  This is especially true when the very small projectile scattering angles are excluded as in the experiment. The black lines in Fig.~\ref{fig:fresco_mdist} show predicted $P(m)$ distributions integrated over all scattering angles, while the red lines are where the smallest scattering angles are excluded, i.e, $\theta_{C.M.}>4^\circ$ which approximately matched the experimental low-angle acceptance of the HiRA array. 
We note that the $P(m)$ distribution must change as one approaches $\theta_{C.M.}$=0$^\circ$  where projection of the projectile-target orbital angular momentum is restricted to zero. In this case, the $m$-state of the projectile is unchanged during the reaction.

The solid lines in Fig.~\ref{fig:fresco_mdist} show results where a particular multipolarity $K$ is specified, i.e. using the lowest possible value consistent with angular momentum and parity conservation. This corresponds to  $K$=1 for 3/2$^+$ and 5/2$^+$ states, $K$=2 for  3/2$^-$, 5/2$^-$, and 7/2$^-$ states, $K$=3 for the 7/2$^+$ states and $K$=4 for 9/2$^-$ states. The results for the rotational model are shown as the long-dashed lines for the 5/2$^-$, 7/2$^-$, and 9/2$^-$ states. Single-particle excitations are shown as the short-dashed lines. Many of these different calculations  overlap and it is difficult to distinguish them in the figure. 

Given the lack of sensitivity of the predictions to the input parameters, we will not give the details of the calculations. For single-particle excitations, we have considered both proton and neutron excitations. For example in excitation of the 5/2$^-$ level, we have considered  the excitation of a $1s_{1/2}$ proton to the $0d_{5/2}$ orbital and the excitation of a $1p_{3/2}$ neutron to the $1p_{1/2}$ orbital. When the optical-model or coupling parameters of a particular calculation are modified, the changes in the predicted $P(m)$ distributions are similar to those seen in Fig.~\ref{fig:fresco_mdist} when changing the reaction mechanism.

Using the $P(m)$ distributions predicted by FRESCO for $\theta_{C.M.}>4^\circ$ we have calculated the angular distributions for the proton decay to the ground and excited  states of $^{12}$N. In many cases, admixtures of different proton $\ell,j$ decays are possible. In such cases, the angular distributions have contributions from each of the $\ell,j$ components plus interference terms.  We note that for an admixture of $s$ and $d$-wave decay, it can only take a few percent of $d$ configurations to produce significant anisotropy in the decay as long as there is significant spin alignment. In these cases, the anisotropy is produced predominantly from the interference term. For an 3/2$^-$ spin assignment, the $m$-state distribution is predicted to be flat in Fig.~\ref{fig:fresco_mdist}(d) and hence the proton angular distribution will be isotropic. For other candidate spin assignments, the amplitudes of the different possible $\ell,j$ components are varied to best reproduce the experimental angular distributions.  We now turn to discuss the fits to each of the experimental angular distributions which are shown in Fig.~\ref{fig:fitMdist}.

\subsubsection{Peak \textit{3}}
\begin{figure}
%sortcode_addback/tree/O13/fitMdist.C
\includegraphics[scale=0.4]{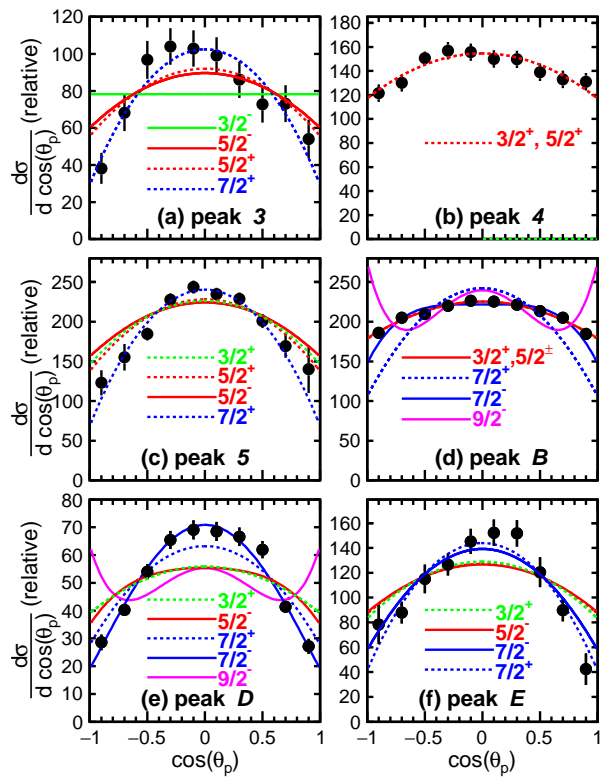}
\caption{Fits to the decay angular distributions with the $m$-state distributions from FRESCO shown in Fig.~\ref{fig:fresco_mdist} for $\theta_{C.M.}>4^\circ$.}
\label{fig:fitMdist}
\end{figure}

As seen in Fig.~\ref{fig:fitMdist}(a), only a spin assignment of 7/2$^+$ allows a good reproduction of the experimental data.  This state proton decays to the 1$^+$ ground state of $^{12}$N by the emission of a $d_{5/2}$ proton and there is only one fit parameter, the overall normalization factor, used to reproduce the data. The fit has a $\chi^2/\nu$=1.24. 

\subsubsection{Peak \textit{4}}
We equate this state to the 4.2(3)-MeV level identified by Suzuki \cite{Suzuki:2012} which was assigned a spin of either 1/2$^-$, 3/2$^-$, 3/2$^+$, or 5/2$^+$.  The first two possibilities are associated with isotropic emission inconsistent with observation, therefore the parity of this state is positive with $J$=3/2 or 5/2. Figure~\ref{fig:fitMdist}(b) shows that identical fits are obtained with these two spins and hence the angular distribution does not provide further restrictions on the spin assignment.

\subsubsection{Peak \textit{5}}
Like peak \textit{3}, only a 7/2$^+$ spin assignment can reproduce the experiment angular distribution  [Fig.~\ref{fig:fitMdist}(c)] so this state also  decays to the 1$^+$ ground-state of $^{12}$N  by $d_{5/2}$ proton emission. This fit has a $\chi^2/\nu$=0.86.

\subsubsection{Peak \textit{B}}
\label{sec:peakB}
Spin assignments of  $J^{\pi}$=3/2$^+$, 5/2$^+$, and 5/2$^-$ produce essentially identical fits in  Fig.~\ref{fig:fitMdist}(d) and best reproduce the data.
These assignments fit the data with $\chi^2/\nu$=0.4 and the next best fit with $J^{\pi}$=7/2$^-$ (solid blue curve) has $\chi^2/\nu$=2.0 and was thus discarded. For a 5/2$^-$ assignment, identical fits can be obtained with different combinations of the $s_{1/2}$, $d_{5/2}$, and $d_{3/2}$ amplitudes.  If we ignore the $d_{3/2}$ component, then the fit indicates predominantly $s$-wave decay with a 16(5)\% $d_{5/2}$ component. However, the $d_{3/2}$ component can also be included with contributions up to 13.6\%.  For the positive-parity spin assignments, the angular distributions were fit with  an admixture of $p_{1/2}$ and $p_{3/2}$ decays with the former restricted  to either 86(4)\% or 18(4)\% for the 5/2$^+$ value  and to either 67(8)\% or less than 2\% for the 3/2$^+$ value.

\subsubsection{Peak \textit{D}}
This state was best fit with a 7/2$^-$ spin assignment giving  $\chi^2/\nu$=0.83. As this states decays to  the 2$^-$ state of $^{12}$N, both $d_{3/2}$ and $d_{5/2}$ proton emissions are possible giving us two fit parameters, i.e. the amplitudes of each component. In the fit, the $d_{3/2}$ component was not tightly restricted but must fall in range from  27\% to 92\%.

\subsubsection{Peak \textit{E}}
Fits to the angular distribution for peak \textit{E} in Fig.~\ref{fig:fitMdist}(f) show only  spin assignments of 7/2$^+$ and 7/2$^-$ are possible with 
$\chi^2/\nu$=1.09 and 1.66, respectively. For a 7/2$^+$ assignment, the fit corresponds to an admixture of $d_{5/2}$ plus $d_{3/2}$ proton emission with the former restricted to the range from 14\% to 63\%. For the 7/2$^-$ assignment, only $p_{3/2}$ proton decay is possible.

\section{NCSMC calculations}
\label{sec:NCSMC}
To study properties of $^{13}$O and its mirror $^{13}$B theoretically, we apply the no-core shell model with continuum (NCSMC)~\cite{Baroni:2013,Baroni:2013a,Navratil:2016} and use chiral nucleon-nucleon (NN) and three-nucleon (3N) interactions as input. We start with the microscopic Hamiltonian
\begin{equation}
H=\frac{1}{A}\sum_{i<j=1}^A\frac{(\vec{p}_i-\vec{p}_j)^2}{2m} + \sum_{i<j=1}^A V^{NN}_{ij} + \sum_{i<j<k=1}^A V^{3N}_{ijk} \, ,\label{H}
\end{equation}
which describes nuclei as systems of $A$ non-relativistic point-like nucleons interacting through realistic inter-nucleon interactions. The modern theory of nuclear forces is based on the framework of chiral effective field theory (EFT)~\cite{Weinberg:1990,Weinberg:1991}. In the present work, we adopt the NN+3N chiral interaction applied in Ref.~\cite{Gysbers:2019}, denoted as NN N$^4$LO+3N(lnl), consisting of an NN interaction up to the fifth order (N$^4$LO) in the chiral expansion~\cite{Entem:2017} and a 3N interaction up to next-to-next-to-leading order (N$^2$LO) using a combination of local and non-local regulators. The 3N low-energy constants (LECs) have been fitted to the $^3$H binding energy and $\beta$-decay half-life. For a faster convergence of our 
calculations with respect to the many-body basis size we softened the chiral interaction through the similarity renormalization group (SRG) technique~\cite{Wegner:1994,Bogner:2007,Roth:2008,Jurgenson:2009}. The SRG unitary transformation induces many-body forces, included here up to the three-body level. The four- and higher-body induced terms are small at the $\lambda_{\mathrm{SRG}}{=}1.8$ fm$^{-1}$ resolution scale used in present calculations.

In the NCSMC, the many-body scattering problem is solved by expanding the wave function on continuous microscopic-cluster states, describing the relative motion between target and projectile nuclei (here $^{12}$N and the proton for $^{13}$O and $^{12}$B and the neutron for $^{13}$B), and discrete square-integrable states, describing the static composite nuclear system (here $^{13}$O and $^{13}$B). The idea behind this generalized expansion is to augment the microscopic cluster model, which enables the correct treatment of the wave function in the asymptotic region, with short-range many-body correlations that are present at small separations, mimicking various deformation effects that might take place during the reaction process. The NCSMC wave function for $^{13}$O is represented as
\begin{align}
\ket{\Psi^{J^\pi T}_{A\texttt{=}13, \frac{3}{2}}} = &  \sum_\lambda c^{J^\pi T}_\lambda \ket{^{13} {\rm O} \, \lambda J^\pi T} \nonumber \\
& +\sum_{\nu}\!\! \int \!\! dr \, r^2 
                 \frac{\gamma^{J^\pi T}_{\nu}(r)}{r}
                 {\mathcal{A}}_\nu \ket{\Phi^{J^\pi T}_{\nu r, \frac{3}{2}}} \, . \label{ncsmc_wf_13O}
\end{align}
The first term of Eq.~(\ref{ncsmc_wf_13O}) consists of an expansion over square-integrable energy eigenstates of the $^{13}$O nucleus indexed by $\lambda$. The second term, corresponding to an expansion over the antisymmetrized channel states in the spirit of the resonating group
method (RGM)~\cite{Wildermuth:1977,Tang:1978}, is given by
\begin{align}
\ket{\Phi^{J^\pi T}_{\nu r, \frac{3}{2}}} = &\Big[ \big( \ket{^{12} {\rm N} \, \lambda_{12} J_{12}^{\pi_{12}}T_{12}} \ket{p \, \tfrac12^{\texttt{+}}\tfrac12} \big)^{(sT)}
Y_\ell(\hat{r}_{12,1}) \Big]^{(J^{\pi}T)}_{\frac{3}{2}} \nonumber\\ 
&\times\,\frac{\delta(r{-}r_{12,1})}{r\;r_{12,1}} \, .
\label{eq_13O_rgm_state}
\end{align}
Here, the index $\nu$ represents all relevant quantum numbers except for those explicitly listed 
on the left-hand side of the equation, and the subscript $\frac{3}{2}$ is the isospin projection, i.e., $(Z-N)/2$. The coordinate $\vec{r}_{12,1}$ in Eq.~(\ref{eq_13O_rgm_state}) is the separation vector between the $^{12}$N target and the proton. The ${\mathcal{A}}_\nu$ in Eq.~(\ref{ncsmc_wf_13O}) antisymmetrizes the projectile proton with the $^{12}$N nucleons. Analogous equations apply for $^{13}$B.

The eigenstates of the aggregate ($\ket{^{13} {\rm O} \, \lambda J^\pi T}$) and target 
($\ket{^{12} {\rm N} \, \lambda_{12} J_{12}^{\pi_{12}}T_{12}}$) nuclei, as well as of $^{13}$B and $^{12}$B, are all obtained by means of the NCSM~\cite{Navratil:2000,Navratil:2000a,Barrett:2013} using a basis of many-body harmonic oscillator wave functions with the same frequency, $\Omega$, and maximum number of particle excitations $N_{\rm max}$ from the lowest Pauli-allowed many-body configuration. In this work we used the harmonic oscillator frequency of $\hbar \Omega = 18$ MeV that minimizes the ground-state energy of the studied nuclei in large $N_{\rm max}$ spaces.

The discrete expansion coefficients  $c_{\lambda}^{J^{\pi}T}$ and the continuous relative-motion amplitudes $\gamma_{\nu}^{J^{\pi}T} (r)$ are the solution of the generalized eigenvalue problem derived by representing the Schr\"{o}dinger equation in the model space of the expansions (\ref{ncsmc_wf_13O})~\cite{Navratil:2016}. The resulting NCSMC equations are solved by means of the coupled-channel $R$-matrix method on a Lagrange mesh~\cite{Descouvemont:2010,Hesse:1998,Hesse:2002}.

In general, the sum over the index $\nu$ in Eq.~(\ref{ncsmc_wf_13O}) includes all the mass partitions involved in the formation of the composite system $^{13}$O, i.e., $^{12}$N+$p$, $p$+$p$+$^{11}$C, $^{10}$C+$^3$He etc. For technical reasons, we limit the present calculations to the $^{12}$N+$p$ clusters of Eq.~(\ref{eq_13O_rgm_state}) and similarly for $^{13}$B to the $^{12}$B+$n$ clusters. In the expansion (\ref{eq_13O_rgm_state}), we include the lowest two positive-parity NCSM eigenstates of $^{12}$N, $1^+$ and $2^+$, and the lowest negative parity eigenstate with $J^\pi{=}2^-$. These states correspond to the lowest three experimental states of $^{12}$N with the $1^+$ the bound ground state and the $2^+$ and $2^-$ narrow resonances. The next experimental resonance, $1^-$, is rather broad and it would be unrealistic to approximate it as a bound NCSM eigenstate. Therefore, we do not include any other $^{12}$N NCSM eigenstates. In the first term of the expansion (\ref{ncsmc_wf_13O}), we include the lowest 12 negative parity and the lowest 15 positive parity NCSM eigenstates of $^{13}$O with $J{=}1/2{-}9/2$ and $T{=}3/2$. This is sufficient to cover the energy region of interest. The same number of NCSM eigenstates is used for $^{12}$B and $^{13}$B. We note that the present NCSMC calculations with chiral NN+3N interactions are the first that include target states of both parities.

We performed NCSMC calculations up to $N_{\rm max}{=}7$ basis space. The use of a still larger space is technically not feasible at present. For example, the basis dimension reaches 115 million in $N_{\rm max}{=}7$ $^{13}$O and $^{13}$B NCSM calculations. In addition to the standard {\it ab initio} NCSMC calculations, we performed the so-called NCSMC-pheno calculations~\cite{DohetEraly:2016,Calci:2016} with the input NCSM excitation energies of $^{12}$N and $^{12}$B adjusted to experimental values. The NCSMC-pheno approach allows one to obtain a more realistic theoretical description of $^{13}$O and $^{13}$B. In particular in the $N_{\rm max}{=}7$ NCSMC-pheno calculation, the $^{12}$N ($^{12}$B) $2^+$ excitation energy was modified from the calculated 0.50 (0.46) MeV to the experimental 0.96 (0.95) MeV and the $2^-$ excitation energy from the calculated 3.82 (3.89) MeV to the experimental 1.19 (1.67) MeV. While the $2^+$ energy modification is modest, the $2^-$ excitation energy shift is more substantial. To get a more realistic $2^-$ energy, one would need a larger $N_{\rm max}$ basis or, better, to include the $p$+$p$+$^{11}$C mass partition in the NCSMC calculations. 

\begin{table}
\caption{Levels prediction from the NCSMC calculations for $^{13}$O and $^{13}$B. The $E_{\rm th}$ is the energy with respect to the $^{12}$N+$p$ and $^{12}$B+$n$ threshold, respectively.}
\label{tbl:NCSMC}
\begin{ruledtabular}
\begin{tabular}{c c c c}
$J^{\pi}$ & $E_{\rm th}$ (MeV) & $E^*$ (MeV)   & $\Gamma$ (MeV) \\
\hline
\multicolumn{4}{c}{$^{13}$O}    \\
  \hline
3/2$^-$    &  -1.70   &   0.00   \\ 
3/2$^+$   &    1.10   &   2.80    &   0.52   \\
1/2$^{+}$ &    1.22   &   2.92    &   1.21   \\
5/2$^+$   &    1.68   &   3.38    &   0.11   \\
3/2$^+$   &    2.36   &   4.06    &   2.09   \\
3/2$^-$    &    2.37   &   4.07    &   1.00   \\
5/2$^+$   &    4.30   &   6.00    &   1.52   \\
7/2$^+$   &    4.78   &   6.48    &   0.96   \\
1/2$^-$    &    5.38   &   7.07    &  0.80    \\
7/2$^+$   &    5.50   &   7.20    &  2.30   \\
9/2$^+$   &    5.65   &   7.35    &  1.48 \\
5/2$^-$    &    5.99   &   7.69    &  1.24  \\ 
\hline 
\multicolumn{4}{c}{$^{13}$B}    \\
  \hline
3/2$^-$   & -4.96   &  0.00   \\   
3/2$^+$  & -0.96   &  4.00  \\
5/2$^+$  & -0.62   &  4.34  \\
1/2$^+$  & -0.41   &  4.55  \\
3/2$^+$  &   0.73   &  5.69   &  0.03  \\
3/2$^-$  &    1.20   &  6.16   &  0.04  \\
\end{tabular}
\end{ruledtabular}
\end{table}
The predicted energies and widths of levels in $^{13}$O and $^{13}$B are listed in Table~\ref{tbl:NCSMC} and compared to the experimental level scheme in Fig.~\ref{fig:levelScheme}. Results from the $N_{\rm max}{=}7$ NCSMC-pheno calculation are presented. For a comparison, the $N_{\rm max}{=}7$ {\it ab initio} NCSMC predicts the $3/2^-$ $^{13}$O ($^{13}$B) ground state energy of -1.74 MeV (-4.99 MeV), which is very close to the NCSMC-pheno results. The experimental value is -1.514 MeV (-4.878 MeV), i.e., a very reasonable agreement. The calculated ground-state energy depends only weakly on the NCSMC basis size. In the $N_{\rm max}{=}5$ space, the NCSMC predicts -1.94 MeV and -5.06 MeV for $^{13}$O and $^{13}$B, respectively. 

\begin{figure}
  \includegraphics[width=0.5\textwidth]{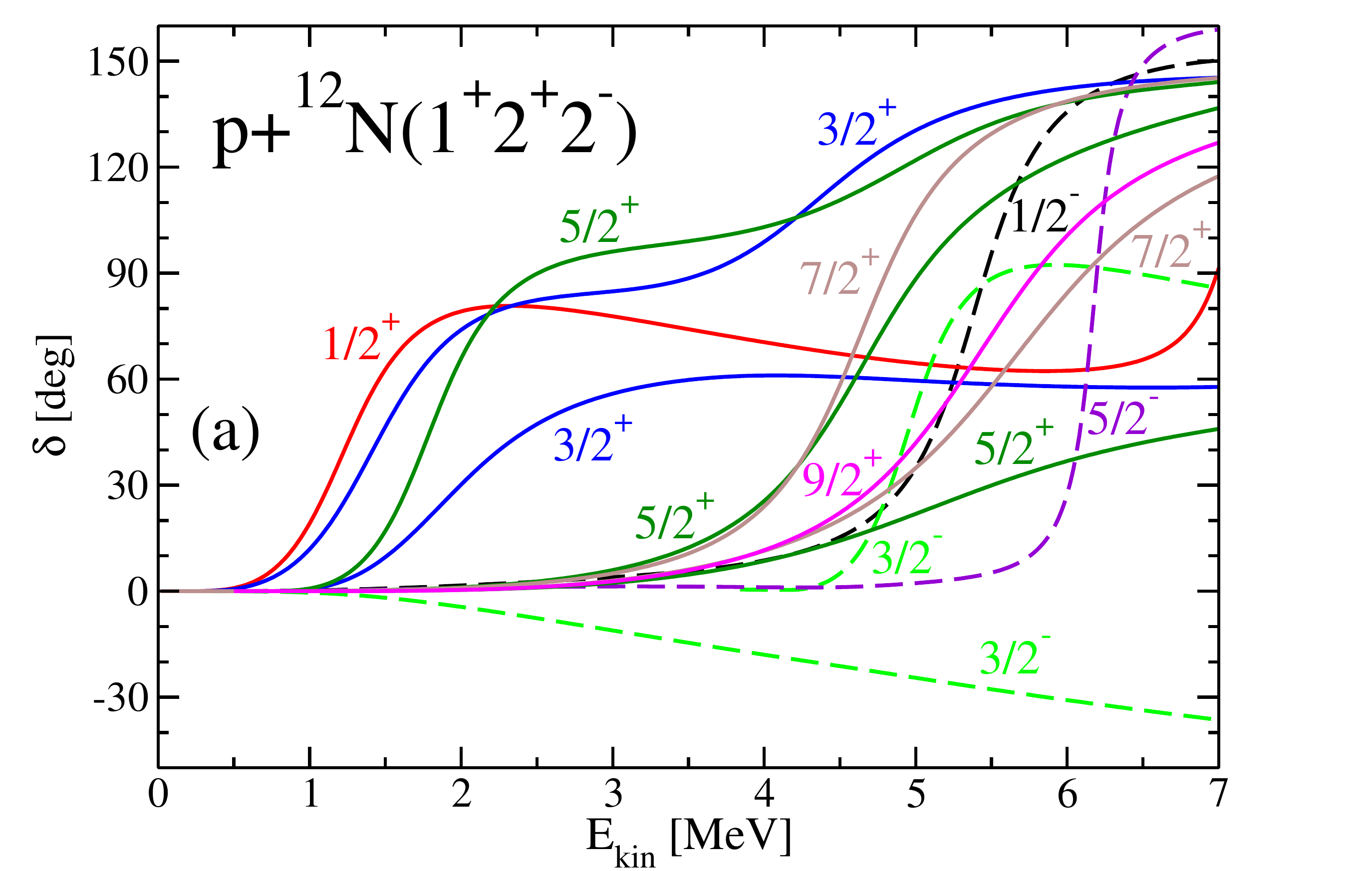}
  \includegraphics[width=0.5\textwidth]{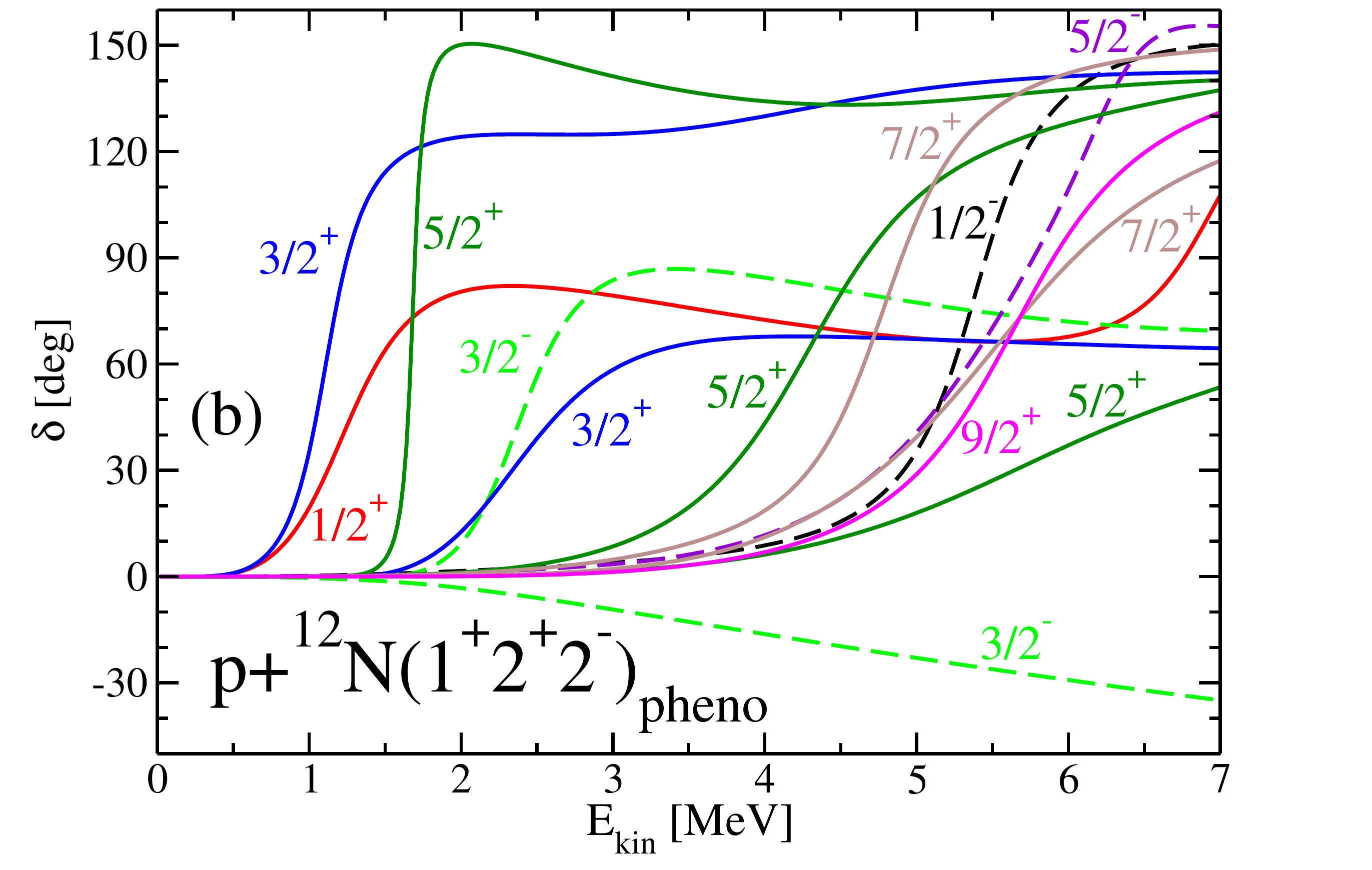}
\caption{The $p$+$^{12}$N eigenphase shift dependence on the energy in the center of mass  $E_{\rm kin}$  which is the same as $E_{\rm th}$ in Table~\ref{tbl:NCSMC} for $E_{\rm th} > 0$. Results obtained in the $N_{\rm max}{=}7$ {\it ab initio} NCSMC (a) and NCSMC-pheno approach (b). The chiral NN N$^4$LO+3N(lnl) interaction was used. See the text for further details.}
\label{fig:NCSMC_eigenphsh}
\end{figure} 
In Fig.~\ref{fig:NCSMC_eigenphsh}, we present selected $p$+$^{12}$N eigenphase shifts obtained in the {\it ab initio} NCSMC and in the NCSMC-pheno calculations. The lowest $1/2^+$ and $3/2^+$ resonances are dominated by the $^2S_{1/2}$ and $^4S_{3/2}$ $^{12}$N$_{1^+}$+$p$ channels, respectively while the lowest $5/2^+$ and the second $3/2^+$ resonances are dominated by the $^6S_{5/2}$ and $^4S_{3/2}$ $^{12}$N$_{2^+}$+$p$ channels, respectively. The $3/2^-$ resonance is dominated by the $^4S_{3/2}$ $^{12}$N$_{2^-}$+$p$ channel. The second $5/2^+$ resonance is a mixture of $^2D_{5/2}$ and $^4D_{5/2}$ $^{12}$N$_{1^+}$+$p$ as well as $^4D_{5/2}$ and $^6D_{5/2}$ $^{12}$N$_{2^+}$+$p$ channels. The $7/2^+_1$ and $9/2^+$ resonances are dominated by $^6D_{7/2}$ and $^6D_{9/2}$ $^{12}$N$_{2^+}$+$p$ channels, respectively, while the $7/2^+_2$ resonance is in the $^4D_{7/2}$ $^{12}$N$_{1^+}$+$p$ channel. The most striking consequences of the phenomenological adjustment of the $^{12}$N excitation energies is the significant decrease of the $5/2^+$ resonance width due to the increase of the $2^+$ excitation energy and the decrease of the $3/2^-$ resonance energy due to the reduction of the $2^-$ excitation energy. We also observe a reduction of the energy and the width of the lowest $3/2^+$ resonance. This is likely a consequence of a smaller mixing between the two $3/2^+$ $s$-waves in the NCSMC-pheno calculation.

For $^{13}$B, we show only the lowest calculated states in Table~\ref{tbl:NCSMC}. The present calculations predict only four bound states while there are seven bound states known experimentally~\cite{Back:2010}.  Our NCSMC calculations could be improved by including the $1^-$ and $0^+$ $^{12}$B states that are bound experimentally. However, such calculations would be technically challenging and since the main focus of the present paper is the structure of $^{13}$O where these states cannot be included, as argued above, we restricted our $^{13}$B calculations to the same model space as that used for $^{13}$O.

\section{DISCUSSION}
\label{sec:discussion} 
The NCSMC predicts the four lowest-energy positive-parity states have configurations which consist predominantly of  a $1s_{1/2}$ proton coupled to the ground or first excited state of $^{12}$N (Sec.~\ref{sec:NCSMC}). The strong $1s_{1/2}$ components lead to large Thomas-Erhman shifts which are conspicuous when comparing the theoretical level schemes of $^{13}$O and $^{13}$B states in Fig.~\ref{fig:levelScheme} and in Table~\ref{tbl:NCSMC}.   Similar configurations are also predicted in shell-model calculations in $psd$ space with the PSDWBT interaction \cite{Warburton:1992}.

 In the experimental data, we find three states (peaks \textit{1}, \textit{2}, and \textit{A}) shifted down below the excitation energy of the lowest-known positive-parity state in $^{13}$B and all were observed to proton decay isotropically consistent with the emissions of $s$-wave protons. The lowest of these (peak \textit{1}) has already been assigned as the $J^{\pi}$=1/2$^+$ state (Sec.~\ref{sec:previous}) and decays to the ground state of $^{12}$N. Its measured decay width of 358(12)~keV is quite similar to a $s$-wave single-particle estimate of 287~keV  obtained with $a_C$=1.4, $a$=0.6 fm, and $r_0$=1.2. The predicted decay width of $\Gamma$=1.21~MeV is larger, but this can be traced to a larger predicted decay energy of 1.22~MeV compared to 0.96~MeV found experimentally, i.e. the increase in the width in the theory is roughly consistent with the single-particle estimate at that decay energy.  Thus this state's dominant configuration is a $1s_{1/2}$ proton coupled to the ground state of $^{12}$N. One also expects a companion 3/2$^+$ level with similar configuration, but different spin coupling. Relative to the energy of the assigned 1/2$^+$ state in $^{12}$B in \cite{Back:2010}, the Thomas-Erhman shift for the 1/2$^+$ state is 1.05~MeV which is comparable to the value of 1.63~MeV predicted by  NCSMC.

Peak \textit{A} proton decays isotropically to the first excited state of $^{12}$N. Its decay width of 54(19) keV is similar to an $s$-wave single-particle estimate of 58~keV. The predicted width is larger at 110~keV, but again this can be traced to the increased decay energy in the theory. Therefore, this  state can  be interpreted as a $1s_{1/2}$ proton coupled to the 2$^+$ second-excited state of $^{12}$N. Its spin is thus either 5/2$^+$ or 3/2$^+$. Assuming is it the 5/2$^+$ value, the 
lowest of these two states in the NCSMC, then taking its analog in $^{13}$B as preferred in \cite{Back:2010}, its Thomas-Erhman shift is 0.63~MeV. This is still significant but smaller than the value of 1.36~MeV predicted in the NCSMC calculations.

The third low-energy  state identified in this work is peak \textit{2} which proton decays isotropically to the ground state of $^{12}$N and is located in between the 1/2$^+$ and 5/2$^+$ levels. We have tentatively assigned this to be the companion 3/2$^+$ state to the 1/2$^+$ state mentioned previously. However, we find it difficult to understand its small decay width. With a decay energy of 1.49~MeV, the estimated $s$-wave single-particle decay width is 1.17~MeV considerably larger than the experimental value of 55(19)~keV.  The predicted width in the NCSMC  is 520~keV which is consistent with a single-particle $s$-wave estimate at the predicted decay energy.  So while its energy is below all known excited states in the mirror nucleus, suggesting it is associated with a Thomas-Erhman shift,  some continuum-coupling effect has reduced its width. We note, that the standard ansatz that the width should be given by the spectroscopic factor times the single-particle value is not  always correct \cite{Charity:2019a}.  

The NCSMC also predicts a fourth positive-parity state with a large Thomas-Erhman shift at $E^*$=4.06~MeV, the 3/2$^+_{2}$ state, with a  configuration largely of an $s$-wave proton coupled to the 2$^+$ $^{12}$N core. In the experimental level scheme, we have two candidate 3/2$^+$ states (peaks \textit{B} and \textit{4}) at similar excitation energies, but neither of these proton decays to the 2$^+$ state of $^{12}$N and thus we find no candidate for this state in the data. Similarly the NCSMC predicts a 3/2$^-$ state at about the same excitation energy whith a large Thomas-Erhman shift and a configuration dominated by $\pi(1s_{1/2})\otimes^{12}$N$_{2^-}$ (Sec.~\ref{sec:NCSMC}). However we again have no experimental candidate for a 3/2$^-$ state which proton decays to the 2$^-$ state of $^{12}$N. Perhaps neither of these states are excited in reaction. We note, however, that the NCSMC $3/2^-$ state might correspond to the 3.29(5)-MeV level reported in Ref.~\cite{Skorodumov:2007}.

Above $E^*\approx4$~MeV, the NCSMC level scheme has a $\approx$2~MeV gap which is not observed in the experimental level scheme. Of the higher NCSMC levels, we find only a few of candidates experimentally. The predicted 7/2$^+_{1}$ state has a dominant configuration of $\pi(0d_{5/3})\otimes^{12}$N$_{2^+}$.  Peak \textit{E} can be a 7/2$^+$ state based on Fig.~\ref{fig:fitMdist}(f) and it also decays to the 2$^+$ state of $^{12}$N. Its excitation energy is only $\approx$0.5~MeV below that of the predicted 7/2$^+$ level  and its width is within 10\% of the predicted value. The observed $7/2^-$ state at $E^* =$ 5.483(5) MeV decaying to the $2^-$ state of $^{12}$N (peak \textit{D}) can be possibly matched with the lowest $7/2^-$ NCSMC resonance that has the $^4D_{7/2}$ - $^6D_{7/2}$ $^{12}$N$_{2^-}$+$p$ structure although it appears at higher energy $E^*\approx$9 MeV and is broader ($\Gamma\sim 2$~MeV).

In addition to having trouble reproducing the higher-energy $^{13}$O levels, the NCSMC also has difficulties with predicting negative-parity excited states of $^{13}$B. The  lowest predicted negative-parity state is at $E^*$=6.2 MeV, but three such states are known experimentally at $E^*\approx4$~MeV. Also, a second 1/2$^+$ state is suggested experimentally at $E^*$=4.829~MeV \cite{Ota:2008,Bedoor:2016}, but no such state is predicted in the NCSMC below 6~MeV. This 1/2$^+$ state may be the mirror of the level associated with peak \textit{C} in this work as it decays isotropically and its excitation energy differs by only $\approx$40~keV. As argued in the previous section, the NCSMC calculations could address some of these issues by including the $1^-$ and $0^+$ $^{12}$B states that are bound experimentally.

Including coupling to the two-body continuum associated with a proton and higher-lying excited states in $^{12}$N in the NCSMC is expected to decrease the excitation energy of the higher-lying states in the mirror pair. This may lead to better agreement with the experimental data.  However for $^{13}$O, the next highest excited state in $^{12}$N is the wide  ($\Gamma$=1.8 MeV) 1$^-$ state, but with such a large width, coupling to this state should be treated via the three-body continuum, which is beyond our present technical capabilities. 
%For $^{13}$B, there are more bound $^{12}$B states that could be included, but such calculations are  very time consuming.

\begin{figure}
%sortcode_addback/tree/O13/AMD.C
\includegraphics[scale=0.43]{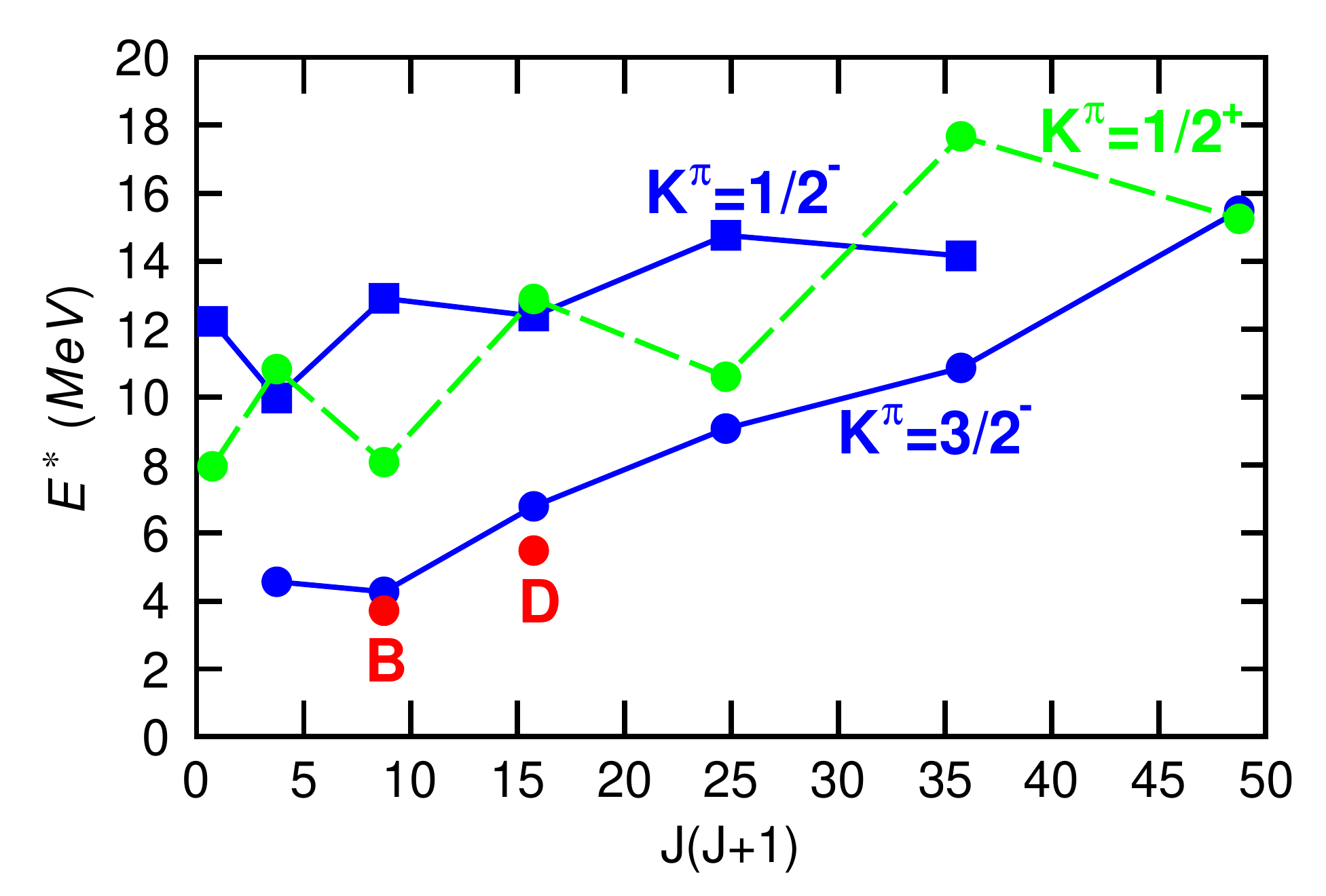}
\caption{Spin dependence of rotational bands associated with deformed cluster configurations in $^{13}$B predicted with the AMD calculations of  Ref.~\cite{Kanada:2008}. The red data points are for peaks \textit{B} and \textit{D} in $^{13}$O, with the former spin assignment taken as 5/2$^-$. }
\label{fig:AMD}
\end{figure}

The presence of low-energy $J$=7/2 states in the experimental level scheme of Fig.~\ref{fig:levelScheme} may be an indicator of collective rotation. Indeed, there is some discussion of deformed states in $^{13}$B \cite{Kanada:2008,Ota:2008}. In particular, antisymmetrized molecular dynamics (AMD) calculations predict three rotational bands associated with cluster states that possess very large deformations. These are predicted to have 2$\hbar\omega$ and 3$\hbar\omega$ configurations. Figure~\ref{fig:AMD} shows the predicted $K^\pi$=1/2$^+$, 1/2$^-$, and 3/2$^-$ rotational bands for interaction A in Ref.~\cite{Kanada:2008}. The red data points show the location of the peaks \textit{D} and \textit{B} with the latter plotted assuming a spin assignment of 5/2$^-$. These points lie within 1~MeV of the predictions for the $K^\pi$=3/2$^-$ band. For the other interaction considered in \cite{Kanada:2008}, the $K^\pi$=3/2$^-$ band shifts up further from the red data points. Both peaks \textit{B} and \textit{D} have large $R$ values, i.e., are predominantly accompanied by a ground-state $^9$Be target fragment. The other member of the group with large $R$ values is peak \textit{3} with $J^\pi$=7/2$^+$ and excitation energy similar to peak \textit{B}. It is possible this level is also associated with collective rotation and/or has cluster structure, however, the positive-parity rotational band predicted with AMD is much higher in excitation energy (Fig.~\ref{fig:AMD}). As we are comparing $^{13}$B predictions with $^{13}$O experimental levels, there may be some modification as the latter are more unbound. Peak \textit{B} has a very small width of 10(19)~keV especially given it decays largely by $s$-wave proton emission for a 5/2$^-$ assignment (Sec.~\ref{sec:peakB}).  This could be explained by the poor overlap of the cluster configuration with the wavefunctions of the low-lying $^{12}$N levels.

\section{CONCLUSIONS}
\label{sec:conclusions}
Excited resonance states in $^{13}$O were produced via the inelastic scattering of  $E/A$=69.5-MeV $^{13}$O beam particles on a $^9$Be target. The resonance states decayed by one and two-proton emission and the decay products were detected in the HiRA $E$-$\Delta E$ array. The states were resolved in the invariant-mass distributions of the $p$+$^{12}$N and 2$p$+$^{11}$C exit channels.

  As in other inelastic-scattering experiments with fast beams \cite{Charity:2015,Hoff:2017,Hoff:2018,Charity:2018}, strongly anisotropic decay angular distributions of the protons were found for some levels when the  target nucleus remained in its ground state. However, some other resonances had isotropic emission patterns. The spin alignment for different resonance spins is expected to be determined from general considerations and independent of reaction mechanisms, i.e. single-particle or collective excitation. This was demonstrated with DWBA calculations and the predicted $m$-state distributions were used to predict the angular distributions for proton decay. From comparison with the experimental distributions, spin assignments of the observed levels  were made.

The experimental level scheme was compared to predictions of the \textit{ab initio} NCSMC. Calculations were performed for $^{13}$O and $^{13}$B considering the $p$+$^{12}$N and $n$+$^{12}$B continuum, respectively, using chiral NN+3N interactions as input. States of both parities of the $^{12}$N and $^{12}$B nuclei were included. The low-energy 1/2$^+$ and 5/2$^+$ states were found to have significant Thomas-Erhman  shifts as predicted by the theory for which $\pi(1s_{1/2})\otimes^{12}$N$_{1^-}$ and $\pi(1s_{1/2})\otimes^{12}$N$_{2^+}$
configurations are deduced. Three other low-energy states 3/2$^+_1$, 3/2$^+_2$ and 3/2$^-_{2}$ are also  predicted to have configurations with a valence $1s_{1/2}$ proton  and large Thomas-Erhman shifts. We find only a candidate for the 3/2$^+_1$ state, though its decay width is much smaller than expected in the theory. A number of higher-energy states observed in this study were not predicted in the NCSMC theory, and some of these may be rotational states build on cluster configurations predicted with  antisymmetrized molecular dynamics (AMD) calculations \cite{Kanada:2008}.

In summary, we have demonstrated the use of spin alignment induced in inelastic scattering of fast beam as a tool to make spin assignments in invariant-mass studies. The technique could also prove useful for excited states that $\gamma$ decay.

\begin{acknowledgments}
This material is based upon work supported by the U.S. Department of Energy, Office of Science, Office of Nuclear Physics under award numbers DE-FG02-87ER-40316, DE-FG02-04ER-41320, DE-SC0014552, DE-SC0013365 (Michigan State University) and the National Science foundation under grant PHY-156556. J.M. was supported by a Department of Energy National Nuclear Security Administration Steward Science Graduate Fellowship under cooperative agreement number DE-NA0002135. Prepared in part by LLNL under Contract DE-AC52-07NA27344. P.N. was supported by the NSERC Grant No. SAPIN-2016-00033. TRIUMF receives federal funding via a contribution agreement with the National Research Council of Canada. Computing support came from an INCITE Award on the Summit supercomputer of the Oak Ridge Leadership Computing Facility (OLCF) at ORNL, from LLNL institutional Computing Grand Challenge program, and from Compute Canada.
\end{acknowledgments}

%\bibliographystyle{aipauth4-1}
%\bibliography{ref}

\end{document}